# Resistance Maintained in Digital Organisms Despite Guanine/Cytosine-Based Fitness Cost and Extended De-Selection: Implications to Microbial Antibiotics Resistance

Clarence FG Castillo[1], Zhu En Chay[2,3] and Maurice HT Ling[3,4]*

[1]School of Information Technology, Republic Polytechnic, Singapore
[2]Department of Accounting, The University of Melbourne, Australia
[3]Colossus Technologies LLP, Singapore
[4]School of BioSciences, The University of Melbourne, Australia

*Corresponding author: Maurice HT Ling, Colossus Technologies LLP, Singapore, E-mail: mauriceling@acm.org

**Abstract**

Antibiotics resistance has caused much complication in the treatment of diseases, where the pathogen is no longer susceptible to specific antibiotics and the use of such antibiotics are no longer effective for treatment. A recent study that utilizes digital organisms suggests that complete elimination of specific antibiotic resistance is unlikely after the disuse of antibiotics, assuming that there are no fitness costs for maintaining resistance once resistance are established. Fitness cost are referred to as reaction to change in environment, where organism improves its' abilities in one area at the expense of the other. Our goal in this study is to use digital organisms to examine the rate of gain and loss of resistance where fitness costs have incurred in maintaining resistance. Our results showed that GC-content based fitness cost during de-selection by removal of antibiotic-induced selective pressure portrayed similar trends in resistance compared to that of no fitness cost, at all stages of initial selection, repeated de-selection and re-introduction of selective pressure. Paired t-tests suggested that prolonged stabilization of resistance after initial loss is not statistically significant for its difference to that of no fitness cost. This suggests that complete elimination of specific antibiotics resistance is unlikely after the disuse of antibiotics despite presence of fitness cost in maintaining antibiotic resistance during the disuse of antibiotics, once a resistant pool of micro-organism has been established.

## Introduction

Antibiotic-resistant bacteria are substantial burden on the human population in terms of both morbidity and mortality, and this drives the need for new discovery and use of antibiotics for treatment. Antibiotic resistance can be naturally occurring. An example of that is that Penicillin-resistant *Bacillus licheniformis* spores were isolated from dried soil from the roots of plants preserved in British Museum since 1689 [1]. Lee et al. [2] suggest that oral antibiotics and food chemicals, such as preservatives, may not be completely absorbed by the gastrointestinal tract, leading to exposure of sub-lethal concentrations to intestinal flora, which may lead to development of tolerance to oral antibiotics and food chemicals [3-5]. This is supported by the observation of the first vancomycin-resistant bacterium, vancomycin-resistant enterococcus in the mid-1980s, after less than a decade of vancomycin use [6]. Once traits conferring resistance are acquired, it can be transmitted to sensitive strains using means such as horizontal gene transfer within intestinal micro-organisms [7].

A number of studies have demonstrated that the maintenance of resistance traits can incur a fitness cost compared to its' antibiotic-sensitive counterparts [8-10] as resistance mutations may disrupt normal physiological processes. The incurrence of fitness cost had been suggested to be able to revert resistant strains back to sensitivity after extended withdrawal of previously resistant antibiotics [11] but with limitations. A few studies suggest that the maintenance of resistance trait in the absence of antibiotics may not incur fitness cost [12-14]. Blot et al. [12] and Dutta [13] have evidence showing that resistant strains may be fitter than the parental sensitive strains. Experimental examination into the gain-loss-re-gain of resistance traits will require sequencing of individual organisms or bacterium, as each organism may be genetically different from each other during the process. Hence, such experimental studies are difficult and expensive.

Computer-simulated evolution of digital organisms (DOs) has been used in many evolutionary studies [15-17] as it offers the ability to examine each genome non-destructively. Castillo and Ling [18] suggests that complete elimination of resistance is unlikely, even after disuse of antibiotics assuming that no fitness cost is needed to maintain antibiotic resistance, when a resistant pool of micro-organisms have been established.

A number of studies suggest that Guanine-Cytosine (commonly known as GC) content is an important fitness trait. Mann and Chen [19] demonstrated that GC content of bacterial genome correlates to their environmental niche. Lightfield et al. [20] suggested that translational efficiency can be constraint by GC content. Povolotskaya et al. [21] examine the prevalence of stop codons in bacteria and found that TAA is preferred for micro-organisms with low GC content while TGA is preferred for micro-organisms with high GC content. This is supported by Bohlin et al. [22] suggesting that selective pressure towards a varying genomic GC contents in micro-organisms are present. The studies show that GC contents are not random but selective. Hershberg and Petrov [23] suggest that deviation from an organism-specific GC contents can incur a fitness cost. Castillo and Ling [18] uses a sequence of 5 consecutive zeros, up to a maximum of 10 blocks, to represent antibiotic resistance. In a binary genome, we can consider '0' to represent guanine and cytosine while '1' to represent adenine and thymine. Given that the initial genome of the DOs is a sequence of 500 bases of alternate binaries, the GC content is 50%. Thus, antibiotic resistance can be considered a high GC trait. In this study, we use 50% GC (equal numbers of 1s and 0s in the genome) as the optimal GC content and increasing deviation from this optimal as fitness cost, to examine the gain-loss-re-gain of resistant traits where maintenance of resistant traits will incur a fitness cost.

In this study, we examine the gain-loss-re-gain of resistant traits where maintenance of resistant traits will incur a fitness cost. The fitness cost can be calculated as deviation from expected GC content. The results from this study will be compared against Castillo and Ling [18]. We hypothesize that incurred fitness cost will result in a greater reduction in population resistance in the absence of antibiotics acting as compared to a case whereby no fitness cost is needed to maintain resistance.

## Materials and Methods

### DOSE platform

The details of DOSE (Digital Organism Simulation Environment) have been described [24-26]. Generally, a population consists of one or more DOs and each DO consists of a genome and a set of status. One or more chromosomes of different lengths make up a genome. Commonly used nucleotides are binary, integer, or alphanumeric. Each chromosome has background mutation rate and an additional mutation rate can be added in the simulation. In addition, defined mutations on a specific base can be made. The genome can be read and executed by Ragaraja interpreter [27]. DO statuses contain information regarding the organism (such as identity and deme), as well as vital conditions (such as age), allowing for traceability across generations. One or more populations can reside in the virtual world, consisting of one or more ecological cells arranged in one or more dimensions. During simulation, 6 simulation-dependent operations are executed on each population. Pre-mating population control was first executed to remove unhealthy organisms. Mutation scheme is performed after first step, using point mutations and gene translocation. Thirdly, fitness is measured before and after mating has occurred. Fourthly, post-mating population control is performed to remove unfit organisms. After the processing is done, ad-hoc generational events - catastrophic events are performed. Lastly, organisms are free to migrate to another ecological cell.

### Simulation setup

The simulation setup is identical to Castillo and Ling [18]. A population of one hundred DOs is contained within one ecological cell used for simulation. The genome of each DO is one chromosome of 500 binary nucleotides. The ancestral chromosome is 500 bases of alternate binaries; that is, $[1010101010]_{50}$. 1% or 5 point mutations per generation form the background mutation.

### Experiment 1 (Initial gain of resistance)

The simulation experiments are based on Castillo and Ling [18]. 25 replicates of 200 generations per replicate were simulated. Pre-determined nucleotide sequence of 11 zeros is representative of antibiotic-resistance. Fitness score for each DO was defined as the number of consecutive zeros within a block for 10 blocks. Three rules were used. Firstly, a minimum of 2 consecutive zeros was required to be considered a block. For example, '101101011' would result in a fitness score of 0 while '101001011' would result in a fitness score of 2 as there were at least 2 consecutive zeros. Secondly, each block contributed 10% of the maximum fitness score. This implies that a maximum of 10 blocks were used to calculate fitness score. All DOs will undergo random mutations and replications in each generation, resulting in 200 DOs. Fitness-proportionate selection (FPS) was used to select 100 DOs. The probability of a DO being selected for the next generation is directly proportional to its fitness score. The fitness score of the selected 100 DOs were recorded.

### Experiment 2 (Loss of resistance)

Each of the 25 populations at 200[th] generation (corresponding to final generation of Experiment 1) was revived and simulated for another 5000 generations to mimic the loss of resistant traits, giving the final generation count of 5200. In each generation, a fitness cost of each DO calculated as the percentage deviation between equal numbers of ones and zeros in the entire genome (representative of 50% GC content). The computation of fitness cost is as follows:

$$\text{Fitness cost (\%)} = \frac{\left[\left(x - \frac{x+y}{2}\right) + \left(y - \frac{x+y}{2}\right)\right]}{x+y}$$

where x is the number of ones and y is the number of zeros. For example, if an organism's genome is made up of 200 ones and 300 zeros, its fitness cost will be 20 or 20% deviation from 50% GC content.

$$\left[\text{Fitness cost (\%)} = \left(\frac{\left(|200-250| + |300-250|\right)}{500}\right)\right]$$

The fitness score is defined as the difference between 100 and the fitness cost. In this case, the fitness score will be 80, which will be used for FPS. Fitness score for each of the 100 surviving DOs were recorded.

Similar to Castillo and Ling [18], we estimated the number of generations needed for the population to revert to pre-selective fitness score, indicated by the average fitness score of un-selected population as control, after the initial decline of fitness score after initial removal of selective pressure. In our previous study [18], fitness score from generation 2000 to 5200 are used to fit a linear regression model, Fitness score = $\beta_1$(Generation) + $\beta_0$. This model can be used to estimate the number of generations needed for the fitness to decline to pre-selection. 95% confidence interval for the number of generations can be estimated by 95% confidence interval of the gradient of decline ($\beta_1$) using Jack-knife re-sampling.

**Experiment 3 (Re-gain of resistance)**

Reviving a population at 5200$^{th}$ generation (corresponding to the last generation of Experiment 2) and continuing the simulation for another 200 generations were used to mimic the re-gain of resistance as a result of re-introduction of antibiotics after extended discontinuation. All processes, including FPS, used in these 200 generations were identical to that of Experiment 1 (initial gain of resistant trait).

**Experiment 4 (Repeated loss and re-gain of resistance)**

Repeated loss and re-gain of antibiotic resistance was carried out to mimic multiple cycles of re-introduction and disuse of antibiotics. Populations at 5200$^{th}$ generation (corresponding to the last generation of Experiment 3) were revived and simulated for 5000 generations of resistance loss using the same procedure as Experiment 2 (loss of resistance). This gain and loss of resistant trait was repeated for another 3 cycles, yielding a total of 26000 generations. The average fitness scores for each re-gain of resistance simulation (5201$^{st}$ to 5400$^{th}$ generation as Gain-2, 10401$^{st}$ to 10600$^{th}$ generation as Gain-3, 15601$^{st}$ to 15800$^{th}$ generation as Gain-4, and 20801$^{st}$ to 21000$^{th}$ generation as Gain-5) were analyzed.

### Results and Discussion

It is postulated that the antibiotic resistant traits may decrease after disuse of the chemical when maintaining resistance traits will incur a fitness cost [8-10]. However, contradictory results have been reported [28]. Dutta [13] and Jechalke et al. [14] suggest that maintaining resistance trait in the absence of antibiotics may not incur fitness cost in certain situations. In fact, resistant strains may be fitter than the parental sensitive strains [12,13].

We use DOs as a tool to study the gain and loss of resistance traits, which is both difficult and expensive to examine experimentally. Castillo and Ling [18] suggested that once a reservoir of resistant micro-organism has been established, complete elimination of specific antibiotics resistance is unlikely despite after extended disuse of antibiotics, assuming no fitness cost for maintaining resistance. Our current study will be compared with Castillo and Ling [18] to study if fitness cost is a factor in maintaining the resistant trait.

Our results show that the average population fitness score increases over 200 generations as a result of selectivity (Figure 1A) and is significantly higher than no selective pressure or control (Paired t-test p-value $< 1 \times 10^{-50}$).

This is supported by recent studies showing direct relationship between emergence of antibiotic resistant strains and amount of prescribed antibiotics [29-33]. The average population fitness score under selective pressure in this study (generation 1 to 200) is not significantly different (paired t-test p-value $> 0.05$) from the corresponding results study by Castillo and Ling [18], where identical selective pressure and simulation setup are used for comparability of results.

All 25 populations were subjected to 5000 generations of selective pressure withdrawal to mimic the disuse of specific antibiotics after increased antibiotic resistance. However, in this study, we added the deviation from an optimal GC content of 50% as fitness cost. We hypothesize that the fitness cost incurred to maintain resistance will reduce fitness compared to no fitness cost incurred in our previous study [18]. Our results from the first loss of resistant trait (experiment 2) show a rapid decline of fitness immediately after withdrawal of antibiotic-induced selective pressure, followed by a plateau of fitness (Figure 1B). This is consistent with studies showing decline of resistance after discontinuation of antibiotics [34,35]. Although the trend of fitness is similar with or without fitness cost by comparing results from this study to Castillo and Ling [18], our results suggest that the incorporation of fitness cost does not significantly reduce average population fitness (paired t-test p-value = 0.013), and the fitness score is significantly higher than that of control (Paired t-test p-value = $4.3 \times 10^{-25}$).

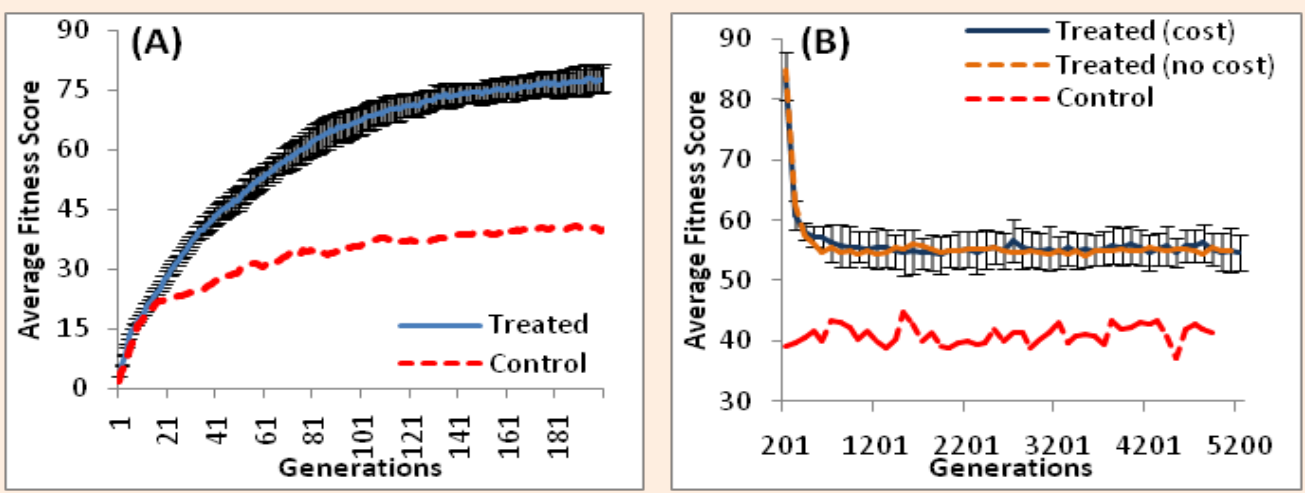

**Figure 1:** Initial gain and loss of resistance trait. Panel A shows the gain of resistance for the first 200 generations while Panel B shows the loss of resistance for the subsequent 5000 generations (corresponding to generation 201 to 5200). Error bars represents standard errors.

Our previous study [18] suggested that the average population fitness is not likely to decline to levels before the use of antibiotics based on regression analyses despite extended de-selection (5000 generations) post-selective pressure. We repeat regression analyses on average population fitness and average top fitness from 2000[th] to 5200[th] generation (Table 1) from this study. This will estimate the number of generations needed for the average population fitness to decline to the level of pre-antibiotic use, which can be estimated by the average population fitness of the control. Our results demonstrate that the rate of fitness change over generations is not statistically significant at 99% confidence, with (this study) or without fitness cost [18]. This implies that the average fitness may remain constant, decrease, or even increase during de-selection. A number of studies had reported similar findings. Enne et al. [36] report an increase of 6.2% in the frequency of sulfonamide-resistant *Escherichia coli* in United Kingdom following 98% decline of sulfonamide prescription. Moreover, there is no further reduction in the prevalence of sulfonamide-resistant *E. coli* 5 years post-study [37]. Nilsson and Laurell [38] report stable proportions of penicillin-resistant *Streptococcus pneumonia* a decade after intervention and reduced penicillin consumption. However, there are also cases of declined antibiotic prevalence. Okon et al. [39] report decline in penicillin, gentamicin, erythromycin, clindamycin, and methicillin resistance rates in sub-Saharan Africa. Lai et al. [40] report decline in cefuroxime, cefotaxime, and cefepime resistance rates in China. These studies suggested that the decline in the occurrence of antibiotic-resistance is not a simple function of stress, but a combination of other factors. Hjalmarsdottir and Kristinsson [41] showed a possibility of redundant resistance traits behaving independently, thus, increasing the unpredictability of antibiotic resistance after antibiotic discontinuation, which is consistent with our results. Although this study only examines GC-content based fitness cost, Hall et al. [42] suggested that fitness cost may be a function of conditional environmental demand on the micro-organism. This means that the role played by fitness costs may not be directly deducible and a possibility of resistant traits being advantageous in certain environmental niches [43] as Miskinyte and Gordo [44] find that 67% of streptomycin-resistant *Escherichia coli* survives better than the susceptible bacteria in the intracellular niche of the phagocytic cells. Baker et al. [45] find that 6 out of 11 fluoroquinolone-resistant *Salmonella typhi* strains to outgrow susceptible strains on competitive growth assays.

Besides the balance of fitness cost incurred and fitness advantage gained from resistance traits, there are studies suggesting that compensatory mutations may occur to circumvent the incurred fitness cost. Albarracin Orio et al. [46] found that the fitness cost incurred by a mutant form of *penicillin-binding protein 2b* (*PBP2b*), giving rise to beta-lactam resistance in *Streptococcus pneumonia*, is fully compensated by acquisition of mutations in *penicillin-binding protein 1a* (*PBP1a*) and *penicillin-binding protein 2x* (*PBP2x*). Zorzet et al. [47] reported that mutations in *accessory gene regulator protein C* (*AgrC*) gene compensates for the fitness cost from *formyl methyl transferase* (*fmt*) gene mutation, which resulted in resistance to high levels of actinonin. Brandis and Hughes [48] report that the most prevalent *bacterial RNA polymerase beta subunit* (*rpoB*) mutation (S531L), resulting in rifampicin resistance in *Mycobacterium tuberculosis* (the causative agent of tuberculosis), can be compensated by mutations in other bacterial RNA polymerase subunits. These suggested alternative means of bypassing the fitness cost incurred by gain of antibiotic resistant traits. This is a possibility which could explain our results do not differ from Castillo and Ling [18] where fitness cost has minimal contribution to reversion of antibiotic resistance.

**Table 1:** Estimated number of generations post-selection that is needed to lose fitness traits. Regression models are generated from gradual loss of fitness after withdrawal of selection pressure (Generation 2000 to 5200) as initial fitness loss (Generation 201 to 1999) may over-estimate the rate of fitness loss.

| | **Study** | **Regression Model** | **Estimated Generations Needed to Lose Fitness** |
|---|---|---|---|
| **Average Top Fitness** | Without fitness cost [18] | (-95% CI) Fitness = 67.3 – 0.000061 Generation | 438,000 |
| | | (Mean) Fitness = 67.3 + 0.000021 Generation | Infinity |
| | | (+95% CI) Fitness = 67.3 + 0.000292 Generation | Infinity |
| | With fitness cost (This study) | (-95% CI) Fitness = 60.7 – 0.000426 Generation | 47,000 |
| | | (Mean) Fitness = 60.7 – 0.000269 Generation | 74,000 |
| | | (+95% CI) Fitness = 60.7 – 0.000100 Generation | 201,000 |
| **Average Population Fitness** | Without fitness cost [18] | (-95% CI) Fitness = 54.7 – 0.000081 Generation | 175,000 |
| | | (Mean) Fitness = 54.7 + 0.000072 Generation | Infinity |
| | | (+95% CI) Fitness = 54.7 + 0.000225 Generation | Infinity |
| | With fitness cost (This study) | (-95% CI) Fitness = 55.1 – 0.000225 Generation | 64,000 |
| | | (Mean) Fitness = 55.1 + 0.000043 Generation | Infinity |
| | | (+95% CI) Fitness = 55.1 + 0.000139 Generation | Infinity |

It is crucial to consider that the acquisition of an initial resistant gene might be a precursor to subsequent resistance. Castillo and Ling [18] suggested that subsequent rate of gain of resistance are faster compared to the rate of initial gain of resistance, despite being interspaced by 5000 generations of resistance decline. Our results showed that the re-gain of resistance (5201st to 5400th generation) after 5000 generations of de-selection is significant faster and higher as shown in Figure 2, (paired t-test p-value = 0.013) than the initial gain of resistance (1st to 200th generation). This is consistent with studies done by Castillo and Ling [18].

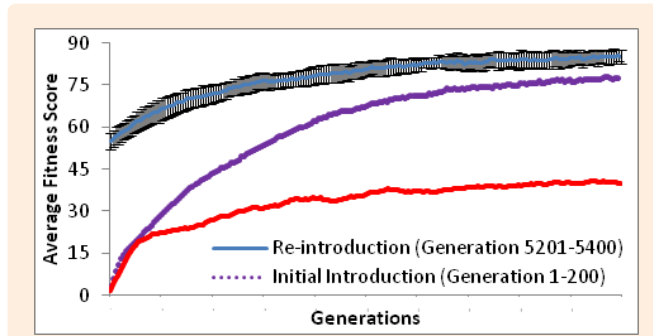

**Figure 2:** Average population fitness score for 200 generations of re-introduction of selective pressure. Re-introduction of selective pressure was carried out from de-selection experiment (generation 5200 in Experiment 3). This is compared to initial introduction of selection to a native population (Experiment 1) and paired t-test is performed between the generation-matched average fitness of initial introduction (generation 1 to 200) and re-introduction (generation 5201 to 5400) is significantly different (Paired t-test p-value = $4.9 \times 10^{-56}$). Error bars represents standard errors.

Our results also suggested that re-introduction of previous antibiotic resistance is likely to result in more rapid resistant formation. However, this has not been shown experimentally. Rapid resistance from antibiotic re-introduction might be due the presence of genetic memory of prior events [49] as suggested by Gajardo and Beardmore [50] to be a species preservation strategy by retaining resistance. Lee et al. [2] suggested that resistance to a specific selective pressure could lead to generic resistance of different selective pressures. A study on single and multi-drug resistant *Mycobacterium smegmatis*. Borrell et al. [51] reports on cases whereby 6 out of 17 isolates that have resistance to both rifampicin and ofloxacin (first and second line antibiotics respectively) show higher fitness costs compared to single antibiotic resistant strains. A study is done by Saddler et al. [52] showed that there are other factors like reproductive numbers and drug efficiency are influential to drug resistance than the cost of resistance fitness.

Our results from the re-gain of resistance (5201st to 5400th generation) simulation show that the average fitness score without fitness cost incurred [18] is significantly higher(Average difference = 0.234Standard error = 0.0378Paired t-test p-value = 3.2 x $10^{-9}$) than the average fitness score with fitness cost incurred (this study). Although this may suggest that fitness cost associated with maintenance of resistance may be acting to revert the population fitness back to susceptibility, our results from subsequent re-gain and loss of selective pressure suggests that such reversion is unlikely as the average population fitness score is substantially higher than that of control population Figure 3 & Table 2 (Paired t-test p-value = $4.3 \times 10^{-25}$). Our results suggest that the fitness cost incurred to maintain antibiotics resistance may have an impact in the level of resistance during discontinuation of antibiotic. Such impact is unlikely to revert resistant strains back to susceptibility as the role of fitness cost is limited for such reversion. Other factors other than fitness cost have to be considered for more comprehensive study [53].

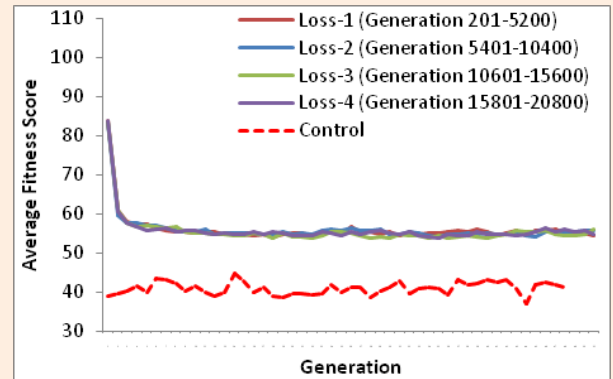

**Figure 3:** Repeated loss of resistance trait. There is no significant difference (Table 2) between consecutive de-selections between with fitness cost (this study) or without fitness cost [18] but there are significant differences in the fitness scores between de-selections and control population (Paired t-test p-value < $2 \times 10^{-25}$, Table 2).

**Table 2:** Paired t-test comparisons of average population fitness between de-selections. Paired t-tests were used instead of one-way ANOVA as the former is targeted towards testing the difference of 2 sets of data; hence, a more appropriate test compared to ANOVA. Our results show that there is no statistical difference between the average population fitness (from 25 replicates) from any 2 de-selections, regardless of selection methods. However, the average population fitness from any de-selection is significantly higher than control.

| Paired t-test Comparisons | p-value |
|---|---|
| Loss-1 (Generation 201 to 5200): With fitness cost (this study) vs without fitness cost [18] | 0.013 |
| Loss-2 (Generation 5401 to 10400): With fitness cost (this study) vs without fitness cost [18] | 0.053 |
| Loss-3 (Generation 10601 to 15400):With fitness cost (this study) vs without fitness cost [18] | 0.406 |
| Loss-4 (Generation 15801 to 20800): With fitness cost (this study) vs without fitness cost [18] | 0.486 |
| Control vs Loss-2 (Generation 5401 to 10400) with fitness cost | 1.3 x $10^{-25}$ |
| Control vs Loss-3 (Generation 10601 to 15400) with fitness cost | 1.6 x $10^{-25}$ |
| Control vs Loss-4 (Generation 15801 to 20800) with fitness cost | 6.2 x $10^{-26}$ |

## Conclusion

In this study, we continue from our previous study [18] to examine the gain-loss-re-gain of resistant traits where fitness cost is incurred to maintain resistant traits, and whether such fitness cost has the potential to revert resistance strains back to susceptibility. Our findings suggest that this is unlikely as the average population fitness in various stages of gain-loss-re-gain of resistant traits with or without fitness cost incurred is not statistically significant. This suggests that fitness cost may have limited role to revert from resistance to susceptibility.